\documentstyle[pra,aps,preprint,floats]{revtex}

\makeatletter
\newcommand{\DoS}{{\mathcal{S}}}
\newcommand{\row}[1]%
{\mathord{\buildrel{\lower3pt%
\hbox{$\scriptscriptstyle\rightarrow$}}\over #1}}
\newcommand{\col}[1]{{#1^{\raisebox{2pt}[\height]%
{$\scriptstyle\downarrow$}}}}
\newcommand{\dyadic}[1]{\mathord{\dyadic@rrow{#1}}}
\newcommand{\dyadic@rrow}[1]{
\begin{picture}(12,12)(-1,0)
\put(0,11){\makebox(0,0)[t]{$\scriptscriptstyle\downarrow$}}
\put(-1,10.5){\makebox(0,0)[l]{$\scriptscriptstyle\longrightarrow$}}
\put(5,0){\makebox(0,0)[b]{$#1$}}
\end{picture}
}
\newcommand{\trans}[1]{{#1}^{\raisebox{0pt}[0pt]{\scriptsize\textrm{T}}}}
\newcommand{\sub}[1]{{\dyadic{#1}}_{\rm sub}}
\newcommand{\tr}[2]{{\mathrm{tr}}_{#1}\left\{#2\right\}}
\newcommand{\Spur}[1]{\mathop{\mathrm{Sp}\left\{#1\right\}}}
\newcommand{\determ}[1]{\mathop{\mathrm{det}\left\{#1\right\}}}
\newcommand{\Rho}{{\mathrm{P}}}
\newcommand{\tRho}{\widetilde{\mathrm{P}}}
\newcommand{\bRho}{\widebar{\mathrm{P}}}
\newcommand{\rktwo}{\Rho_{\rm{rk2}}}
\newcommand{\Bell}{\Rho_{\rm{Bell}}}
\newcommand{\Wern}{\Rho_{\rm{W}}}
\newcommand{\WernA}{\Rho_{\rm{W,1st}}}
\newcommand{\WernB}{\Rho_{\rm{W,2nd}}}
\newcommand{\chaos}{\Rho_{\rm{chaos}}}
\newcommand{\pure}{\Rho_{\rm{pure}}}
\newcommand{\sep}{\Rho_{\rm sep}}
\newcommand{\pureopt}{\Rho^{\rm(opt)}_{\rm{pure}}}
\newcommand{\sepopt}{\Rho^{\rm(opt)}_{\rm{sep}}}
\newcommand{\widebar}[1]{\overline{#1}}
\newcommand{\expect}[1]{\left\langle #1 \right\rangle}
\newcommand{\olexpect}[1]{\bigl\langle #1 \bigr\rangle}
\newcommand{\magn}[1]{%
{\mathchoice{\magn@D{#1}}{\magn@T{#1}}{\magn@S{#1}}{\magn@SS{#1}}}}
\newlength{\@xx}\newlength{\@yy}\newlength{\@zz}
\newcommand{\magn@D}[1]{%
\settoheight{\@xx}{$\displaystyle\left|#1\right|$}%
\settodepth{\@yy}{$\displaystyle\left|#1\right|$}%
\addtolength{\@xx}{\@yy}%
{\,\rule[-\@yy]{\@zz}{\@xx}\,#1\,\rule[-\@yy]{\@zz}{\@xx}\,}}
\newcommand{\magn@T}[1]{%
\settoheight{\@xx}{$\textstyle\left|#1\right|$}%
\settodepth{\@yy}{$\textstyle\left|#1\right|$}%
\addtolength{\@xx}{\@yy}%
{\,\rule[-\@yy]{\@zz}{\@xx}\,#1\,\rule[-\@yy]{\@zz}{\@xx}\,}}
\newcommand{\magn@S}[1]{%
\settoheight{\@xx}{$\scriptstyle\left|#1\right|$}%
\settodepth{\@yy}{$\scriptstyle\left|#1\right|$}%
\addtolength{\@xx}{\@yy}%
{\,\rule[-\@yy]{\@zz}{\@xx}\,#1\,\rule[-\@yy]{\@zz}{\@xx}\,}}
\newcommand{\magn@SS}[1]{%
\settoheight{\@xx}{$\scriptscriptstyle\left|#1\right|$}%
\settodepth{\@yy}{$\scriptscriptstyle\left|#1\right|$}%
\addtolength{\@xx}{\@yy}%
{\,\rule[-\@yy]{\@zz}{\@xx}\,#1\,\rule[-\@yy]{\@zz}{\@xx}\,}}
\newcommand{\olmagn}[1]{\magn{\raisebox{0pt}[0pt][0pt]{$#1$}}}

\makeatother

\tighten

\begin{document}

\title{Separability of entangled q-bit pairs}
\author{Berthold-Georg Englert$^{\ast\dagger}$  
and Nasser Metwally$^{\ddagger}$}
\address{%
$^\ast$Max-Planck-Institut f\"ur Quantenoptik, 
Hans-Kopfermann-Strasse 1, 85748 Garching, Germany\protect\\
$^\dagger$Abteilung f\"ur Quantenphysik, Universit\"at Ulm,
Albert-Einstein-Allee~1, 89081 Ulm, Germany\protect\\
$^\ddagger$Sektion Physik, Universit\"at M\"unchen, 
Theresienstra\ss{}e 37, 80333 M\"unchen, Germany}
\date{18 December 1999; 25 January 2000}

\maketitle
\begin{abstract}%
The state of an entangled q-bit pair is specified 
by 15 numerical parameters that are naturally regarded
as the components of two 3-vectors and a $3\times3$-dyadic. 
There are easy-to-use criteria to check whether a given pair of 3-vectors
plus a dyadic specify a 2--q-bit state; and if they do, whether the
state is entangled; and if it is, whether it is a separable state.
Some progress has been made in the search for analytical
expressions for the degree of separability. 
We report, in particular, the answer in the case of 
vanishing 3-vectors.\\[0.5\baselineskip]
\end{abstract}

\renewcommand{\thefootnote}{(\alph{footnote})}

\section{Introduction}\label{sec:intro}
A q-bit is, in general terms, a binary quantum alternative, for which there
are many different physical realizations.
Familiar examples include the binary alternatives 
of a Stern-Gerlach experiment (``spin up'' or ``spin down''); 
of a photon's helicity (``left handed'' or ``right handed''); 
of two-level atoms (``in the upper state'' or ``in the lower one'');
of Young's double-slit set-up (``through this slit'' or 
``through that slit''); 
of Mach-Zehnder interferometers (``reflected at the entry beam splitter'' or
``transmitted at it''); 
and of Ramsey interferometers (``transition in the first zone'' or 
``in the second zone'').

The actual physical nature of the q-bits in question is irrelevant, however,
for the issues dealt with in this paper.
We are remarking on entangled states of two q-bits, and as far as the somewhat
abstract mathematical properties are concerned, all q-bits are equal.
In particular, the two q-bits in question could be of quite different kinds,
one the spin-$\frac12$ degree of freedom of a silver atom, say, the other a
photon's helicity.
It is even possible, and of experimental relevance 
\cite{DNRa,DNRb,KSE,SKE,BBER},
that both q-bits are carried by the same physical object: the which-way
alternative of an atom (photon, neutron, \dots) passing through an 
interferometer could represent one q-bit, for instance, while its polarization
(or another internal degree of freedom) is the other.

Entangled q-bit pairs are the basic vehicle of proposed quantum communication
schemes, envisioned quantum computers, and the like.
Accordingly, a thorough understanding of the 2--q-bit states they can be in is
highly desirable.

Whereas the possible states of a single q-bit are easily classified with the
aid of a 3-vector (the Bloch vector in one physical  context, the Poincar\'e
vector in another, and analogs of both in general --- we shall speak of Pauli
vectors), the classification of the states of entangled q-bit pairs has not
been fully achieved as yet. 
The obvious reason is the richness of the state space, which is parameterized by
two 3-vectors, one for each q-bit, and a $3\times3$-dyadic that represents 
expectation values of joint observables, so that 15 real numbers are necessary
to specify an arbitrary 2--q-bit state.
A first important division is the one into entangled states and disentangled
ones; a second distinguishes entangled states that are separable from the
non-separable ones (technical definitions are given in Sec.\ \ref{sec:notation}
below). 
The latter ones differ from each other by various properties. 
Among them is the \emph{degree of separability}, which we would like
to express in terms of the said 15 parameters (or rather of the 9 relevant
ones among them, see Sec.\ \ref{sec:invariants}).

In the present paper, which is a progress report in spirit, we'll be content
with an exposition of the formalism we employ and a concise presentation of
some results of particular interest.
A more technical account will be given elsewhere \cite{E+Mprep}.

\section{Notation, terminology, and other preparatory remarks}
\label{sec:notation}
Analogs of Pauli's spin operators are, as usual, used for the description of
the individual q-bits: the set $\sigma_x$, $\sigma_y$, $\sigma_z$ for the
first q-bit, and $\tau_x$, $\tau_y$, $\tau_z$ for the second.
Upon introducing corresponding sets of unit vectors --- $\row{e_x}$,
$\row{e_y}$, $\row{e_z}$ and $\row{n_x}$, $\row{n_y}$, $\row{n_z}$,
respectively, each set orthonormal and right-handed --- we form the vector
operators
\begin{equation}\label{eq:Pauli-ops}
\row{\sigma}=\sum_{\alpha=x,y,z}\sigma_{\alpha}\row{e_{\alpha}}\,,\qquad
\row{\tau}=\sum_{\beta=x,y,z}\tau_{\beta}\row{n_{\beta}}\,.
\end{equation}
We emphasize that the two three-dimensional vector spaces thus introduced are
unrelated and they may have nothing to do with the physical space.
Even if the q-bits should consist of the spin-$\frac12$ degrees of freedom of
two electrons, say, so that an identification with the physical space would be
natural, we could still define the $x$, $y$, and $z$ directions independently
for both q-bits.

Book keeping is made considerably easier if one distinguishes row vectors from
column vectors, related to each other by transposition.
We write
\begin{equation}\label{eq:row-col}
\col{\sigma}=\trans{\row{\sigma}}\,,\qquad
\row{\tau}=\trans{\col{\tau}}\,,\qquad\mbox{et cetera}  
\end{equation}
with a self-explaining notation. 
Scalar products, such as $\olexpect{\row{\sigma}}\cdot\col{\sigma}$ and
$\row{\tau}\cdot\olexpect{\col{\tau}}$ involve a row and a column of the same
type; products of the ``column times row'' kind are dyadics, for which
\begin{equation}\label{eq:dyad-ex}
\col{\sigma}\row{\tau}=
\sum_{\alpha,\beta=x,y,z}
\col{e_{\alpha}}\sigma_{\alpha}\tau_{\beta}\row{n_{\beta}}
\end{equation}
is an important example; it is a column of $e$-type combined with a row of
$n$-type. 

The statistical operators, the \emph{states} for short, for the two q-bits
themselves are given by
\begin{equation}\label{eq:rho1+2}
\rho_1=\frac{1}{2}\left(1+\row{s}\cdot\col{\sigma}\right)\,,\qquad
\rho_2=\frac{1}{2}\left(1+\row{\tau}\cdot\col{t}\right)
\end{equation}
with
\begin{equation}\label{eq:Pauli}
\row{s}=\expect{\row{\sigma}}\,,\qquad \col{t}=\expect{\col{\tau}}\,,
\end{equation}
respectively.
An arbitrary joint 2--q-bit state,
\begin{equation}\label{eq:Rho}
\Rho=\frac{1}{4}\left(1+\row{s}\cdot\col{\sigma}+\row{\tau}\cdot\col{t}
+\row{\sigma}\cdot\dyadic{C}\cdot\col{\tau}\right)  
\end{equation}
involves the \emph{cross dyadic}
\begin{equation}\label{eq:cross}
  \dyadic{C}=\expect{\col{\sigma}\row{\tau}}
\end{equation}
in addition to the \emph{Pauli vectors} $\row{s}$ and $\col{t}$.
The 15 expectation values that constitute $\row{s}$, $\col{t}$, and
$\dyadic{C}$ can be obtained by measuring 5 well chosen 2--q-bit observables,
such as the ones specified in Table \ref{tbl:5obs}.
These 5 observables are pairwise complementary and thus represent
an optimal set in the sense of Wootters and Fields \cite{WF}.
Or, as Brukner and Zeilinger would put it, the left column of Table
\ref{tbl:5obs} lists ``a complete set of five pairs of complementary
propositions''~\cite{BZ}.

\begin{table}[!t]
\caption[Aa]{\label{tbl:5obs}%
A minimal set of five 2--q-bit observables whose measurement supplies all 15
parameters that characterize the state $\Rho$ of Eq.\ (\ref{eq:Rho}).
\rule[-12pt]{0pt}{5pt}}
\begin{tabular}{cc}
\multicolumn{2}{c}{The observable}\\
which identifies the joint eigenstates of
&
determines the three expectation values\\
\hline
$\sigma_x$ and $\tau_x$ &
$\olexpect{\sigma_x}$, $\olexpect{\tau_x}$, $\olexpect{\sigma_x\tau_x}$ \\
$\sigma_y$ and $\tau_y$ &
$\olexpect{\sigma_y}$, $\olexpect{\tau_y}$, $\olexpect{\sigma_y\tau_y}$ \\
$\sigma_z$ and $\tau_z$ &
$\olexpect{\sigma_z}$, $\olexpect{\tau_z}$, $\olexpect{\sigma_z\tau_z}$ \\
$\sigma_x\tau_y$ and $\sigma_y\tau_z$ &
$\olexpect{\sigma_x\tau_y}$, $\olexpect{\sigma_y\tau_z}$, 
$\olexpect{\sigma_z\tau_x}$  \\
$\sigma_y\tau_x$ and $\sigma_z\tau_y$ &
$\olexpect{\sigma_y\tau_x}$, $\olexpect{\sigma_z\tau_y}$, 
$\olexpect{\sigma_x\tau_z}$ 
\end{tabular}
\end{table}

Partial traces,
\begin{equation}\label{eq:part-tr}
\rho_1=\tr{2}{\Rho}\,,\qquad\rho_2=\tr{1}{\Rho}
\end{equation}
extract $\rho_1$ and $\rho_2$ from $\Rho$, of course.
The difference between the product state $\rho_1\rho_2$ and the actual one,
\begin{equation}\label{eq:Edyad}
\Rho-\rho_1\rho_2=\frac{1}{4}\row{\sigma}\cdot\dyadic{E}\cdot\col{\tau}\,,  
\end{equation}
involves the \emph{entanglement dyadic} $\dyadic{E}$, given by
\begin{equation}\label{eq:Edyad-expl}
\dyadic{E}=\dyadic{C}-\col{s}\row{t}=\expect{\col{\sigma}\row{\tau}}
-\expect{\col{\sigma}}\expect{\row{\tau}}\,.
\end{equation}
The state $\Rho$ is \emph{entangled} if ${\dyadic{E}\neq0}$. 

An entangled state $\Rho$ can be a mixture of disentangled ones,
\begin{equation}\label{eq:mix}
\Rho=\sum_kw_k\frac{1}{2}\left(1+\row{s_k}\cdot\col{\sigma}\right)
\frac{1}{2}\left(1+\row{\tau}\cdot\col{t_k}\right)
\qquad\mbox{with $w_k>0$ and $\displaystyle\sum_kw_k=1$,}
\end{equation}
in which case it is \emph{separable}.
The correlations associated with an entangled, but separable state are not of
a quantum nature and can be understood classically.

According to the findings of Lewenstein and Sanpera \cite{LS}, any
2--q-bit state $\Rho$ can be written as a mixture of a separable state $\sep$
and a non-separable pure state $\pure$ ${[\,=\pure^2\,]}$,
\begin{equation}\label{eq:LS}
 \Rho=\lambda\sep+(1-\lambda)\pure\qquad\mbox{with ${0\leq\lambda\leq1}$.} 
\end{equation}
As a rule, there are many different such \emph{LS decompositions} with varying
values of $\lambda$.
Among them is the unique optimal decomposition, the one with the largest
$\lambda$ value,
\begin{equation}\label{eq:LSopt}
\Rho=\DoS\sepopt+(1-\DoS)\pureopt\,,
\end{equation}
where
\begin{equation}\label{eq:DoS}
\DoS=\max\{\lambda\}  
\end{equation}
is the \emph{degree of separability} possessed by $\Rho$; the value $\DoS=0$
obtains only if $\Rho$ itself is a non-separable pure state. 
The number $\DoS$ measures to which extent the correlations associated with 
$\Rho$ are classical; in rough terms, a state $\Rho$ is the more useful for
quantum communication purposes, the smaller its degree of separability.

Therefore, we would like to express $\DoS$ and $\pureopt$ in terms of the
Pauli vectors $\row{s}$, $\col{t}$ and the cross dyadic $\dyadic{C}$ that
specify the state $\Rho$.
We are still searching for the general answer, but for a number of important
special cases the problem is solved already.
We report some of this partial progress below.

Whereas it is relatively easy to find LS decompositions for a given state
$\Rho$, it is usually rather difficult to check whether a certain
decomposition is the optimal one.
Here is what's involved (for ${\lambda>0}$):
\begin{equation}\label{eq:LSpairing}
\parbox{0.7\columnwidth}{%
If $\Rho=\lambda\sep+(1-\lambda)\pure$ is the optimal decomposition, then\\ 
\begin{tabular}{@{\qquad}rp{0.5\columnwidth}@{}}
(a)& {\raggedright%
the state $(1+\varepsilon)^{-1}\left(\sep+\varepsilon\pure\right)$\\ 
is non-separable for $\varepsilon>0$;}\\
and (b)& {\raggedright%
the state $\sep+(1/\lambda-1)\left(\pure-\pure'\right)$\\ 
is either non-positive or non-separable\\ for each $\pure'\neq\pure$.}
\end{tabular}}
\end{equation}
Only $\sep$ and $\pure$ of the optimal decomposition (when ${\lambda=\DoS}$)
meet both criteria.
Unfortunately, their verification is rather complicated even in seemingly
simple cases.

Since the infinitesimal neighborhood of $\pure$ is critical in
(\ref{eq:LSpairing}), the actual value of $1/\lambda-1$ is irrelevant and, as
a consequence, we note an important pairing property:
\begin{equation}\label{eq:pair2}
\parbox{0.7\columnwidth}{%
If $\Rho_{\lambda}=\lambda\sep+(1-\lambda)\pure$ is the optimal LS
decomposition for one value of $\lambda$ in the range $0<\lambda<1$,
then it is optimal also for all other $\lambda$ values.}
\end{equation}
Obviously, a systematic method for identifying all $\sep$s that pair with a
given $\pure$, or vice versa, would  be quite helpful, but we are not aware of
one presently.

\section{Invariants and inequalities}\label{sec:invariants}
The freedom to choose $\col{e_x}$, $\col{e_y}$, $\col{e_z}$ and $\row{n_x}$,
$\row{n_y}$, $\row{n_z}$ to our liking means that unitary transformations that
affect only $\col{\sigma}$, or only $\row{\tau}$, or both separately, turn a
given $\Rho$ into a physically equivalent one.
In terms of the Pauli vectors and the cross dyadic, such \emph{local}
transformations are of the form
\begin{equation}\label{eq:local}
  \col{s}\to\dyadic{O}_{\rm ee}\cdot\col{s}\,,\qquad
\row{t}\to\row{t}\cdot\dyadic{O}_{\rm nn}\,,\qquad
\dyadic{C}\to\dyadic{O}_{\rm ee}\cdot\dyadic{C}\cdot\dyadic{O}_{\rm nn}\,,
\end{equation}
where
\begin{equation}\label{eq:Oee+Onn}
\dyadic{O}_{\rm ee}=\col{e_1}\row{e_x}+\col{e_2}\row{e_y}+\col{e_3}\row{e_z}\,,
\qquad
\dyadic{O}_{\rm nn}=\col{n_x}\row{n_1}+\col{n_y}\row{n_2}+\col{n_z}\row{n_3} 
\end{equation}
are orthogonal unimodular dyadics that relate the $x,y,z$ description to the
$1,2,3$ one.
Since each of them needs 3 parameters for its specification, there must be
${9=15-(3+3)}$ independent combinations of $\row{s}$, $\col{t}$, and
$\dyadic{C}$ that are invariant under (\ref{eq:local}).
These are\footnote{\label{fn:Spur}%
We write $\Spur{\ }$ for the trace of a dyadic in order to avoid confusion
with quantum mechanical traces such as 
$C_{xy}=\olexpect{\sigma_x\tau_y}=\tr{1\&2}{\sigma_x\tau_y\Rho}$.} 
\begin{equation}\label{eq:loc-invs}
\begin{array}{rcl@{\quad}rcl}
a^{(2)}_1&=&\Spur{\trans{\dyadic{C}}\cdot\dyadic{C}}\,, &
a^{(2)}_2&=&\row{s}\cdot\col{s}\,,\qquad
a^{(2)}_3=\row{t}\cdot\col{t}\,,  \\[2ex]
a^{(3)}_1&=&\determ{\dyadic{C}}\,, &
a^{(3)}_2&=&\row{s}\cdot\dyadic{C}\cdot\col{t}\,,   \\[2ex]
a^{(4)}_1&=&\Spur{\left(\trans{\dyadic{C}}\cdot\dyadic{C}\right)^2}\,, &
a^{(4)}_2&=&\row{s}\cdot\sub{C}\cdot\col{t}\,,  \\
a^{(4)}_3&=&\row{s}\cdot\dyadic{C}\cdot\trans{\dyadic{C}}\cdot\col{s}\,, &
a^{(4)}_4&=&\row{t}\cdot\trans{\dyadic{C}}\cdot\dyadic{C}\cdot\col{t}\,, 
\end{array}
\end{equation}
where the dyadic $\sub{C}$ consists of the subdeterminants of $\dyadic{C}$.
All other local invariants can be expressed in terms of the nine $a^{(n)}_m$s.
Important examples are the determinant of the entanglement dyadic,
\begin{equation}\label{eq:detE}
\determ{\dyadic{E}}=\determ{\dyadic{C}}-\row{s}\cdot\sub{C}\cdot\col{t}
=a^{(3)}_1-a^{(4)}_2\,, 
\end{equation}
and the trace of the modulus of the cross dyadic,
\begin{equation}\label{eq:Sp-modC}
\Spur{\magn{\dyadic{C}}}
=\Spur{\left(\trans{\dyadic{C}}\cdot\dyadic{C}\right)^{1/2}}
=\sqrt{\zeta_1}+\sqrt{\zeta_2}+\sqrt{\zeta_3}\,,  
\end{equation}
where $\zeta_1$, $\zeta_2$, and $\zeta_3$ are the three roots of the cubic
equation 
\begin{equation}\label{eq:cubic}
\zeta^3-a^{(2)}_1\zeta^2+\frac{1}{2}\left[\left(a^{(2)}_1\right)^2
-a^{(4)}_1\right]\zeta-\left(a^{(3)}_1\right)^2=0\,.
\end{equation}

Admixing the totally chaotic state $\chaos=\frac{1}{4}$ to the given $\Rho$,
\begin{equation}\label{eq:Rhox}
\Rho_x=(1-x)\chaos+x\Rho\qquad\mbox{with ${0\leq x\leq1}$}\,,  
\end{equation}
amounts to
\begin{equation}\label{eq:Rhox'}
\row{s}\to x\row{s}\,,\qquad\col{t}\to x\col{t}\,,
\qquad\dyadic{C}\to x\dyadic{C}\,.  
\end{equation}
The resulting scaling of the local invariants is
\begin{equation}\label{eq:invs-scal}
  a^{(n)}_m\to x^n a^{(n)}_m\,,
\end{equation}
which is the reason for the grouping in (\ref{eq:loc-invs}).

In addition to the local transformations (\ref{eq:local}), there are also the
\emph{global} ones that represent arbitrary unitary transformations of the
state $\Rho$.
Except for the eigenvalues of $\Rho$, nothing is left unchanged.
In view of the restriction $\tr{1\&2}{\Rho}=1$, there must be 3 global
invariants.
A convenient choice is
\begin{equation}\label{eq:As}
\begin{array}{rcl}
A_2&=&2\left(a^{(2)}_1+a^{(2)}_2+a^{(2)}_3\right)\,,\\[2ex]
A_1&=&8\left(a^{(3)}_2-a^{(3)}_1\right)\,,\\[2ex]
A_0&=&\left(a^{(2)}_1\right)^2-2a^{(2)}_1\left(a^{(2)}_2+a^{(2)}_3\right)
-\left(a^{(2)}_2-a^{(2)}_3\right)^2 \\ &&
-2a^{(4)}_1-8a^{(4)}_2+4a^{(4)}_3+4a^{(4)}_4\,,
\end{array}
\end{equation}
which scale in accordance with $A_k\to x^{4-k}A_k$ under (\ref{eq:Rhox'}).
The $A_k$s are significant because they are the coefficients in the quartic
equation
\begin{equation}\label{eq:quartic}
\kappa^4-A_2\kappa^2+A_1\kappa-A_0=0  
\end{equation}
that determines the eigenvalues of $\Rho$: If $\kappa$ is a solution of
(\ref{eq:quartic}), then ${(1-\kappa)/4}$ is an eigenvalue of $\Rho$.
The absence of the cubic term reflects the unit trace of $\Rho$.

Since $\Rho$ is hermitian, all roots of  (\ref{eq:quartic}) are real by
construction, and ${\Rho\geq0}$ implies the inequalities
\begin{equation}\label{eq:Rho>0}
A_2-A_1+A_0\leq1\,,\qquad 2A_2-A_1\leq4\,,\qquad A_2\leq6\,.  
\end{equation}
They enable one to check whether a given set of $\row{s}$, $\col{t}$,
$\dyadic{C}$ actually defines a state $\Rho$.

The global reflection
\begin{equation}\label{eq:glob-refl}
\row{s}\to-\row{s}\,,\qquad\col{t}\to-\col{t}\,,\qquad\dyadic{C}\to\dyadic{C}  
\end{equation}
has no effect on the local invariants (\ref{eq:loc-invs}), and therefore%
\footnote{\label{fn:barP}%
In the studies by Hill and Wootters \cite{HW,Woo} of what they call
``entanglement of formation'' the state $\bRho$ plays a central role; in
particular the eigenvalues of $\olmagn{\sqrt{\bRho}\sqrt{\Rho}}$ are of interest.}
\begin{equation}\label{eq:barRho}
\bRho=\frac{1}{4}\left(1-\row{s}\cdot\col{\sigma}-\row{\tau}\cdot\col{t}
+\row{\sigma}\cdot\dyadic{C}\cdot\col{\tau}\right)  
\end{equation}
has the same eigenvalues as $\Rho$ and also the same degree of separability
$\DoS$.
Mixtures of both,
\begin{equation}\label{eq:Rhoy}
\Rho_y=\frac{1+y}{2}\Rho+\frac{1-y}{2}\bRho
=\frac{1}{4}\left(1+y\row{s}\cdot\col{\sigma}+y\row{\tau}\cdot\col{t}
+\row{\sigma}\cdot\dyadic{C}\cdot\col{\tau}\right)\,,
\end{equation}
(with ${-1\leq y\leq1}$) have degrees of separability $\DoS_y$ that cannot be
less than that of $\Rho$ and $\bRho$,
\begin{equation}\label{eq:Sy}
\DoS_y\geq\DoS\,,
\end{equation}
which is a useful piece of information because everything is known for the
${y=0}$ case, see Sec.\ \ref{ssec:genWern1} below.

The partial reflection
\begin{equation}\label{eq:part-refl}
\row{s}\to-\row{s}\,,\qquad\col{t}\to\col{t}\,,\qquad\dyadic{C}\to-\dyadic{C}  
\end{equation}
is a non-unitary transformation of $\Rho$, which is turned into
\begin{equation}\label{eq:tRho}
\tRho=\frac{1}{4}\left(1-\row{s}\cdot\col{\sigma}+\row{\tau}\cdot\col{t}
-\row{\sigma}\cdot\dyadic{C}\cdot\col{\tau}\right)\,. 
\end{equation}
Peres \cite{Per} observed that ${\tRho\geq0}$ if $\Rho$ is separable, and his
conjecture that $\Rho$ is separable if ${\tRho\geq0}$ was proven by M., P.,
and R. Horodecki \cite{MPRHor}:
\begin{equation}\label{eq:P3H}
\parbox{0.6\columnwidth}{%
A 2--q-bit state $\Rho$ is separable if its $\tRho$ is non-negative, \\ 
and only then.
}
\end{equation}
Now, since (\ref{eq:part-refl}) affects only two of the nine local invariants
(\ref{eq:loc-invs}), namely $a^{(3)}_1$ and $a^{(4)}_2$ whose sign changes,
the positivity conditions (\ref{eq:Rho>0}) are immediately translated into
corresponding conditions for $\tRho$, and we arrive at this statement:
\begin{equation}\label{eq:tRho>0}
\parbox{0.5\columnwidth}{\raggedright%
If\qquad\ $A_2-A_1+A_0\leq1+16\determ{\dyadic{E}}$\\
and\qquad\  $2A_2-A_1\leq4+16\determ{\dyadic{C}}$\\
then $\Rho$ is separable; if one of the inequalities\\ is violated, then $\Rho$
is not separable.}
\end{equation}
It is therefore a straightforward matter to check whether a certain $\Rho$ is
separable (${\DoS=1}$) or not (${\DoS<1}$).

With the aid of a local transformation (\ref{eq:local}), one can bring a given
$\Rho$ into a generic form.
A standard one refers to the bases for which the cross dyadic is diagonal,
\begin{equation}\label{eq:diagC}
\dyadic{C}=\sum_{\alpha,\beta=x,y,z}
\col{e_{\alpha}}C_{\alpha\beta}\row{n_{\beta}}
=\pm\sum_{k=1}^3\col{e_k}c_k\row{n_k}\qquad
\mbox{for}\ \left\{
\begin{array}{l}
\determ{\dyadic{C}}\geq0\,,\\[1ex]\determ{\dyadic{C}}<0\,,
\end{array}\right.
\end{equation}
where the $c_k$s are the square roots of the $\zeta_k$s in (\ref{eq:Sp-modC}),
ordered in accordance with 
\begin{equation}\label{eq:charvals}
  c_1\geq c_2\geq c_3\geq0
\end{equation}
by convention.
Then, the moduli $\olmagn{\dyadic{C}}$ and $\olmagn{\trans{\dyadic{C}}}$ of
$\dyadic{C}$ and $\trans{\dyadic{C}}$ as well as $\sub{C}$ have simple
appearances, too,
\begin{equation}\label{eq:mod+sub}
\begin{array}{c}\displaystyle
\magn{\dyadic{C}}=\sum_{k=1}^3\col{n_k}c_k\row{n_k}\,,\qquad
\magn{\trans{\dyadic{C}}}=\sum_{k=1}^3\col{e_k}c_k\row{e_k}\,,\\[2ex]
\sub{C}=\col{e_1}c_2c_3\row{n_1}+\col{e_2}c_3c_1\row{n_2}
+\col{e_3}c_1c_2\row{n_3}\,,
\end{array}
\end{equation}
so that
\begin{equation}\label{eq:C=O*mod}
\dyadic{C}=\pm\dyadic{O}_{\rm en}\cdot\magn{\dyadic{C}}
=\pm\magn{\trans{\dyadic{C}}}\cdot\dyadic{O}_{\rm en}\,,  
\end{equation}
where
\begin{equation}\label{eq:Oen}
 \dyadic{O}_{\rm en}=\sum_{k=1}^3\col{e_k}\row{n_k} 
\end{equation}
is an orthogonal unimodular dyadic.

For example, a pure state $\pure$ has $A_2-A_1+A_0=1$, $2A_2-A_1=4$, $A_2=6$
and its generic form is  
\begin{equation}\label{eq:generic-pure}
\pure=\frac{1}{4}\left(1+p\sigma_1-p\tau_1
-\sigma_1\tau_1-q\sigma_2\tau_2  -q\sigma_3\tau_3\right)
\end{equation}
with
\begin{equation}\label{eq:sigma_k,tau_k}
\sigma_k=\row{\sigma}\cdot\col{e_k}\,,\qquad
\tau_k=\row{n_k}\cdot\col{\tau}\qquad
\mbox{for ${k=1,2,3}$}  
\end{equation}
and ${0\leq p\leq1}$, ${q\equiv\sqrt{1-p^2}}$.
Thus, up to local transformations, pure states are characterized by a single
parameter, namely the common length of the Pauli vectors,
${p=(\row{s}\cdot\col{s})^{1/2}}{=(\row{t}\cdot\col{t})^{1/2}}$.
A pure state is separable if ${p=1}$, not separable if ${p<1}$.
For ${p=0}$, one has the so-called Bell states
\begin{equation}\label{eq:Bell}
\Bell=\frac{1}{4}
\left(1-\row{\sigma}\cdot\dyadic{O}_{\rm en}\cdot\col{\tau}\right)  
\end{equation}
with $\dyadic{O}_{\rm en}$ as in (\ref{eq:Oen}).

\section{Special cases}\label{sec:special}
\subsection{Werner states}\label{ssec:Wern}
The so-called Werner states \cite{Wern} are (pseudo-)mixtures of Bell states
and the chaotic state,
\begin{equation}\label{eq:Wern0}
\Wern=(1-x)\chaos+x\Bell=\frac{1}{4}
\left(1-x\row{\sigma}\cdot\dyadic{O}_{\rm en}\cdot\col{\tau}\right)\,,
\end{equation}
where $\Wern\geq0$ requires $-\frac{1}{3}\leq x\leq1$ since the eigenvalues
of $\Wern$ are $\frac{1}{4}(1+3x)$ and $\frac{1}{4}(1-x)$, the latter being
three-fold. 
Here one has
\begin{equation}\label{eq:Wern1}
\row{s}=0\,,\qquad\col{t}=0\,,\qquad\dyadic{C}=-x\dyadic{O}_{\rm en}  
\end{equation}
and finds%
\footnote{\label{fn:LSnumerics}%
The numerical findings of Lewenstein and Sanpera \cite{LS} agree well with
this analytical result.}
\begin{equation}\label{eq:Wern2}
\DoS=\left\{\begin{array}{c@{\quad\mbox{if}\quad}l}
1& -\frac{1}{3}\leq x \leq\frac{1}{3}\,,\\[1ex]
\frac{3}{2}(1-x) & \frac{1}{3}< x \leq1\,,
\end{array}\right.
\end{equation}
for the degree of separability.
The pure state of the optimal LS decomposition is the Bell state that appears
in (\ref{eq:Wern0}).

\subsection{Generalized Werner states of the first kind}\label{ssec:genWern1}
States $\Rho$ for which
\begin{equation}\label{eq:1stWern0}
\row{s}=0\,,\qquad\col{t}=0\,,\qquad
\dyadic{C}=\pm\dyadic{O}_{\rm en}\cdot\magn{\dyadic{C}}\quad\mbox{arbitrary}    
\end{equation}
represent a first generalization of the Werner states (\ref{eq:Wern0}).
The $y=0$ states of (\ref{eq:Rhoy}) are among them.

The eigenvalues of
$\row{\sigma}\cdot\dyadic{O}_{\rm en}\cdot\magn{\dyadic{C}}\cdot\col{\tau}$ are
${c_1+c_2-c_3}$, ${c_1-c_2+c_3}$, ${-c_1+c_2+c_3}$, and ${-c_1-c_2-c_3}$ with
the $c_k$s as in (\ref{eq:diagC}), and the positivity of
\begin{equation}\label{eq:1stWern1}
  \WernA=\frac{1}{4}
\left(1\pm\row{\sigma}\cdot\dyadic{O}_{\rm en}\cdot\magn{\dyadic{C}}
\cdot\col{\tau}\right)
\end{equation}
then requires that the triplet $(c_1,c_2,c_3)$ --- which is not a 3-vector ---
is inside the tetrahedron that R. and M. Horodecki speak of in Ref.\
\cite{RMHor}. 

The degree of separability of a state $\WernA$ is given by
\begin{equation}\label{eq:1stWern2}
\DoS=\left\{\begin{array}{c@{\quad\mbox{if}\quad}lcl}
1& \determ{\dyadic{C}}\geq0 &\mbox{or}& \Spur{\magn{\dyadic{C}}}\leq1\,,\\[1ex]
\frac{3}{2}-\frac{1}{2}\Spur{\magn{\dyadic{C}}} &
 \determ{\dyadic{C}}<0 &\mbox{and}& \Spur{\magn{\dyadic{C}}}>1\,,
\end{array}\right.
\end{equation}
and the pure state of the optimal LS decomposition is the Bell state
(\ref{eq:Bell}) with $\dyadic{O}_{\rm en}$ from (\ref{eq:1stWern0}).

\subsection{Generalized Werner states of the second kind}\label{ssec:genWern2}
A second generalization of the Werner states is obtained by replacing the Bell
state in (\ref{eq:Wern0}) by an arbitrary pure state with $0<p,q<1$ in
(\ref{eq:generic-pure}). 
Then one has
\begin{equation}\label{eq:2ndWern0}
\WernB=\frac{1+3x}{4}\pure+\frac{1-x}{4}\left(1-\pure\right)\,.  
\end{equation}
Upon denoting by $q_0$ the $q$ parameter of the pure state in the optimal LS
decomposition, one gets
\begin{equation}\label{eq:2ndWern1}
\DoS=\left\{\begin{array}{c@{\quad\mbox{if}\quad}l}
1& -\frac{1}{3}\leq x \leq (1+2q)^{-1}\,,\\[1ex]
\displaystyle
1-\frac{(1+2q)x-1}{2q_0} & (1+2q)^{-1}< x \leq1\,,
\end{array}\right.
\end{equation}
and $q_0=\sqrt{1-p_0^2}$ is the largest value that obeys
\begin{equation}\label{eq:2ndWern2}
\frac{1+x-2xpp_0}{q_0}\leq\left(qx-\frac{1-x}{2}\right)
+\left(qx-\frac{1-x}{2}\right)^{-1}(x-x^2p^2)\,.
\end{equation}
This gives ${q_0>q}$ for ${x<1}$ and ${q_0\to q}$ in the limit $x\to1$; 
the extreme value $q_0=1$ is reached if $x$ is in the range
\begin{equation}\label{eq:2ndWern3}
  \frac{1}{1+2q}<x\leq\frac{3/4}{q-1/4+\sqrt{(1-q)(1+q/2)}}\,,
\end{equation}
and then a Bell state shows up in the optimal LS decomposition.

\subsection{States of rank 2}\label{ssec:rank2}
A state $\Rho$, for which $A_2-A_1+A_0=1$ and $2A_2-A_1=4$, has eigenvalues
$0$ (two-fold), $(1+x)/2$, and $(1-x)/2$ with $x^2=(A_2-2)/4\leq1$. 
For $x^2<1$, such a $\Rho$ is of rank 2.
Its generic form is
\begin{equation}\label{eq:rk2-0}
\rktwo=\frac{1}{2}\left(\Sigma_0+x_1\Sigma_1+x_2\Sigma_2+x_3\Sigma_3\right)
\qquad\mbox{with $x_1^2+x_2^2+x_3^2=x^2$}\,,
\end{equation}
where
\begin{equation}\label{eq:rk2-1}
\Sigma_0=\frac{1}{2}\left(1
+\sigma_3\cos\gamma_1\cos\gamma_2
+\tau_3\sin\gamma_1\sin\gamma_2
+\sigma_1\tau_1\sin\gamma_1\cos\gamma_2
+\sigma_2\tau_2\cos\gamma_1\sin\gamma_2\right)
\end{equation}
projects onto the two-dimensional subspace in question.
By convention, the parameters $\gamma_1$ and $\gamma_2$ are such that
${\pi/2\geq\gamma_1\geq\gamma_2\geq0}$. 
They also appear in the expressions for $\Sigma_{1,2,3}$,
\begin{equation}\label{eq:rk2-2}
\begin{array}{rcl}
\Sigma_1&=&\displaystyle\frac{1}{2}\left(
\sigma_1\sin\gamma_1
+\tau_1\cos\gamma_2
+\sigma_1\tau_3\sin\gamma_2
+\sigma_3\tau_1\cos\gamma_1\right)\,,\\[1ex]
\Sigma_2&=&\displaystyle\frac{1}{2}\left(
\sigma_2\sin\gamma_2
+\tau_2\cos\gamma_1
+\sigma_2\tau_3\sin\gamma_1
+\sigma_3\tau_2\cos\gamma_2\right)\,,\\[1ex]
\Sigma_3&=&\displaystyle\frac{1}{2}\bigl(
\sigma_3\sin\gamma_1\sin\gamma_2
+\tau_3\cos\gamma_1\cos\gamma_2\\ 
&&\displaystyle\phantom{\frac{1}{2}\bigl(}
-\sigma_1\tau_1\cos\gamma_1\sin\gamma_2
-\sigma_2\tau_2\sin\gamma_1\cos\gamma_2
+\sigma_3\tau_3\bigr)\,,
\end{array}
\end{equation}
which are analogs of Pauli's spin operators for the subspace defined by
$\Sigma_0$.
Their basic algebraic properties are
\begin{equation}\label{eq:rk2-3}
\begin{array}{c}\displaystyle
\Sigma_0\Sigma_k=\Sigma_k\qquad\mbox{for ${k=0,1,2,3}$}\,,\\[1ex]
\displaystyle
\Sigma_j\Sigma_k=\delta_{jk}\Sigma_0+i\sum_{l=1}^3\epsilon_{jkl}\Sigma_l
\qquad\mbox{for ${j,k=1,2,3}$}\,.
\end{array}
\end{equation}

The pure rank-2 states (\ref{eq:rk2-0}) have $x_1^2+x_2^2+x_3^2=1$. 
If $\sin\gamma_1\cos\gamma_2=0$, which is to say that 
${\pi/2=\gamma_1=\gamma_2}$ or ${\gamma_1=\gamma_2=0}$, 
then all the states (\ref{eq:rk2-0}) are separable; 
otherwise the separable ones have $x_2=0$,
$x_3=\tan\gamma_2/\tan\gamma_1\equiv\cos(2\vartheta)$ and
$\magn{x_1}\leq\sin(2\vartheta)$ with $0\leq\vartheta\leq\pi/4$.
For $\gamma_1>\gamma_2$ there are two separable pure states, for
$\pi/2>\gamma_1=\gamma_2>0$ (and thus $\vartheta=0$) there is only one.
Equivalent observations about rank-2 states have been made by Sanpera, Tarrach,
and Vidal \cite{STV}.

For $\sin\gamma_1\cos\gamma_2>0$, the pairing of (\ref{eq:LSpairing}) and
(\ref{eq:pair2}) leads to pairs of three different kinds, viz.\
\begin{equation}\label{eq:rk2-4}
\begin{tabular}{rl}
(a)& $\pure$ with $x_1=0$ \& $\sep$ with $\magn{x_1}<\sin(2\vartheta)$\,,\\
(b)& $\pure$ with $x_1\geq0$ \&  $\sep$ with $x_1=\sin(2\vartheta)$\,,\\
(c)& $\pure$ with $x_1\leq0$ \&  $\sep$ with $x_1=-\sin(2\vartheta)$\,.
\end{tabular}
\end{equation}
For a given rank-2 state (\ref{eq:rk2-0}) this means the following.
If the inequality
\begin{equation}\label{eq:rk2-5}
\left[(1+x_3)\sin\vartheta-\magn{x_1}\cos\vartheta\right]  
\left[(1-x_3)\cos\vartheta-\magn{x_1}\sin\vartheta\right]
\leq x_2^2\sin\vartheta\cos\vartheta  
\end{equation}
holds, then 
\begin{equation}\label{eq:rk2-6}
\DoS=\frac{(1-x^2)/2}{1-x_3\cos(2\vartheta)-\magn{x_1}\sin(2\vartheta)}
\end{equation}
and the pairs (\ref{eq:rk2-4})(b) or (\ref{eq:rk2-4})(c) apply for ${x_1>0}$ and
${x_1<0}$, respectively.
If (\ref{eq:rk2-5}) is violated, then the optimal LS decomposition involves
pair (\ref{eq:rk2-4})(a) and
\begin{equation}\label{eq:rk2-7}
\DoS=\frac{1}{\sin^2(2\vartheta)}\left(1-x_3\cos(2\vartheta)
-\sqrt{[x_3-\cos(2\vartheta)]^2+[x_2\sin(2\vartheta)]^2}\,\right)  
\end{equation}
is the degree of separability.

\section{Outlook}\label{sec:outlook}
Since any arbitrary 2--q-bit state $\Rho$ is a mixture of two rank-2 states, the
complete solution of the rank-2 case can be used in an iterative manner to
arrive at LS decompositions of a given $\Rho$. 
It is hoped that the optimal decomposition can be found this way, and we shall
report results in due course.

\section{Acknowledgments}
We are grateful for very helpful discussions with H.-J. Briegel and I. Cirac.
BGE would like to thank H. Rauch and his collaborators at the Atominstitut in
Vienna, where part of this work was done, for the hospitality they
provided, and the Technical University of Vienna for financial support.
BGE would also like to thank M. O. Scully and the physics faculty at Texas
A\&M University, where another part of this work was done, for their
hospitality and financial support.  
NM would like to thank the Egyptian government for granting a fellowship.

\vfill

\setlength{\fboxsep}{10pt}
\begin{center}
\framebox{\parbox{0.75\columnwidth}{%
\begin{center}
This paper has been submitted to\\
Journal of Modern Optics\\      
as a contribution to the\\ 
Proceedings of the Workshop on Entanglement and Decoherence,\\ 
held at Gargnano/Italy, 20-25 September 1999.
\end{center}}}      
\end{center}

\vfill


\begin{references}
\bibitem{DNRa}
S. D\"urr, T. Nonn, and G. Rempe,
\nat \textbf{395}, 33 (1998).
\bibitem{DNRb}
S. D\"urr, T. Nonn, and G. Rempe,
\prl \textbf{81}, 5705 (1998).
\bibitem{KSE}
P. G. Kwiat, P. D. D. Schwindt, and B.-G. Englert,
in \textit{Mysteries, Puzzles, and Paradoxes in Quantum Mechanics},
edited by R. Bonifacio (AIP CP461, 1999).
\bibitem{SKE}
P. D. D. Schwindt, P. G. Kwiat, and B.-G. Englert,
\pra \textbf{60}, 4285 (1999).
\bibitem{BBER}
G. Badurek, R. J. Buchelt, B.-G. Englert, and H. Rauch,
Nucl.\ Instrum.\ Meth.\ A, in print.
\bibitem{E+Mprep}
B.-G. Englert and N. Metwally, in preparation.
\bibitem{WF}
W. K. Wootters and B. D. Fields, 
Ann.\ Phys.\ (NY) \textbf{191}, 363 (1989).
\bibitem{BZ}
\v{C}.\ Brukner and A. Zeilinger,
\prl \textbf{83}, 3345 (1999).
\bibitem{LS}
M. Lewenstein and A. Sanpera,
\prl \textbf{80}, 2261 (1998).
\bibitem{HW}
S. Hill and W. K. Wootters,
\prl \textbf{78}, 5022 (1997).
\bibitem{Woo}
W. K. Wootters,
\prl \textbf{80}, 2245 (1998).
\bibitem{Per}
A. Peres,
\prl \textbf{77}, 1413 (1996).
\bibitem{MPRHor}
M. Horodecki, P. Horodecki, R. Horodecki,
\pl \textbf{A223}, 1 (1996). 
\bibitem{Wern}
R. F. Werner, 
\pra \textbf{40}, 4277 (1989).
\bibitem{RMHor}
R. Horodecki and M. Horodecki,
\pra \textbf{54}, 1838 (1996).
\bibitem{STV}
A. Sanpera, R. Tarrach, and G. Vidal,
\pra \textbf{58}, 826 (1998).
\end{references}
\end{document}